1.11.2003.

# Autodetaching States of Atomic Negative Ions.

V.A.Esaulov*
Laboratoire des Collisions Atomiques et Moléculaires
CNRS-Unversité Paris Sud, bât 351, Université Paris Sud, Orsay, France
vladimir.esaulov@u-psud.fr, ORCID: 0000-0002-7263-9685.

A critical review on excited states of negative ions has been given recently by Buckman and Clark (1994), in a major update of the review by Schulz (1973). However, for completeness we include a summary of some known (mainly low lying) excited states. We include a brief discussion and indicate some problematic points with additional references. The observation of resonances in gas phase atomic collisions is discussed in this context. Finally we mention some very nice new results concerning the observation of these resonant states in particle surface interactions and briefly discuss the role that they can play. A table of electron affinities of atomic negative ions is provided at the end.

## 1. Resonant states of negative ions.

In most cases there do not exist stable singly excited states of negative ions such as e.g. 1snl for $H^-$. There exist however a large number of short lived doubly excited states (e.g. $2s^2$ or 2s2p for $H^-$) of negative ions lying above the detachment limit. Autodetaching states were first observed in electron scattering studies and have more recently been investigated in photodetachment measurements and in heavy particle collisions. They have been frequently subdivided into two classes (see Taylor et al.1966 and Schulz 1973). The first class called *Feshbach resonant states* corresponds to the possibility of an excited atom forming a negative ion by electron attachment with an overall reduction of energy (both electrons being in excited orbitals). The electron affinity of the excited state is normally less than the excitation energy (generally the affinities are in the 0 to 0.5 eV range). Because decay into the parent state is impossible, decay occurs into the ground or excited states.

$$A^{-*} \rightarrow A\ (A^*) + e^-.$$

This requires a change in configuration and Feshbach resonances are relatively long lived ($10^{-14}$s). The second class of resonances called *shape resonances* occur when the interaction potential has a sufficiently large barrier caused by the centrifugal force which can temporarily trap the electron in the well. Clearly shape resonances are to be expected for $l \neq 0$ waves. Shape resonances usually decay into their parent states and hence their lifetimes are shorter than for Feshbach states. This "traditional" classification scheme is not ideal and other classification schemes have been proposed. Buckman and Clark (1994) have preferred to use a scheme developed by Read and coworkers (Brunt et al 1976) for inert gas resonances. For a detailed discussion of these classifications the reader is referred to this review.

When an autodetaching state of a negative ion lies above an excited state of the neutral atom, decay can occur either to the ground state or to an excited state. Thus resonances below the n = 3 level of hydrogen may decay to both the n = 2 or n = 1 states of hydrogen. An important question then, is one concerning the branching ratio for these decay channels. Macek and Burke (1967) have for instance calculated the relative transition probabilities for several H states below the n = 3 level. It was found that decay into the n = 2 level is predominant. An interesting situation may exist in certain cases as e.g. for $F^-$ and other halogen anions, where decay of an autodetaching state corresponding to a given core configuration $2s^22p^4(^1D)3s^2\ ^1D$ could occur into an excited state of another configuration (in the above case the $(2s^22p^4(^3P)3p)\ ^2P$ or $^2D$ states). The branching ratio of decay into these states and into the ground state of e.g. $F(2s^22p^5)$ is not known. These and other results are illustrated in the next paragraph.

Turning again to the question of singly excited states we note that in certain cases (e.g. for metal atoms like Hg, Mg, Li, etc.) (see e.g. Johnston and Burrow 1982), there may exist single particle shape resonances lying just above the detachment limit. These are of the e.g. (nsnp $^3$P) type in case of $Li^-$, and are attributed to the polarisation effect of an electron on the target resulting, for $l \neq 0$, in the creation of a centrifugal barrier in which the electron is trapped. Note that the question of the various (unstable) states of negative ions corresponding to the ground electronic configuration (e.g. $N^{-*}\ 2p^4\ ^1D$ and $^1S$) has been discussed extensively by Bunge in chapter1.

Negative ion resonant states have been studied theoretically for a number of negative ions. A detailed discussion can be found in the review of Buckman and Clark (1994). From a practical point of view, it seems interesting to point out here that reasonable estimates of the positions of negative ion resonances may be obtained using a modification of the Rydberg formula (Read 1977). This modified Rydberg formula has been found to give good estimates (to an accuracy of 100 meV or better) of the energies of negative ion resonances of the inert gas atoms, oxygen, halogen negative ions etc...

We recall that for states having a (core)nl configuration the energy is given by,

$$E_{nl} = I \quad \frac{RZ^2}{(n \quad _{nl})^2}$$

where I is the ionization energy, R is the Rydberg energy, Z is the effective charge of the core ( $Z = Z_{nucl} - N + 1$ for a N electron atom), n and l are the radial and angular quantum numbers and $\delta_{nl}$ is the quantum defect, which represents a measure of the penetration of the outer nl electron into the core and of the difference between the field it experiences there and the Coulomb field of a point charge Z. Read (1977) suggests that for configurations of the type (core)nl-nl' (where l = l' or l≠l' and the core contains no other nl or nl' electrons) the energy of the configuration can be parametrised by the modified Rydberg formula,

$$E_{n\ln l'} = I \quad \frac{R(Z \quad )^2}{(n \quad _{nl})^2} \quad \frac{R(Z \quad )^2}{(n \quad _{nl'})^2}$$

where I is the energy of the core configuration, Z is the charge of the core ($N_{nucl} - N + 2$, for a N electron atom) and $\sigma$ is an effective screening constant which depends on the degree of correlation of the two Rydberg electrons. The quantum defects used in this formula should be rendered free of effects of magnetic interaction and exchange correlations.

For the case of configurations of the type (core)$np^N$ with N=1-6, assuming that each electron produces an effective screening $\sigma$ on each of the other (N - 1 ) electrons, and that each electron experiences an effective core field that can be parametrised by the quantum defect $\delta_{nl}$ of a single np electron in the atom having the configuration (core)np, one obtains for the energy,

$$E(np^N) \quad E(np^0) \quad \frac{NR(Z \quad (N \quad 1) \quad )^2}{(n \quad _{np})^2}$$

Average values of the screening constant $\sigma$ for some of the studied cases are listed below.

| Class | Term | $\sigma$ |
|---|---|---|
| a | $ns^2\,{}^1S$ | 0.25 |
| b | $nsnp\,{}^3P°$ | 0.241 |
| c | $nsnp\,{}^1P°$ | 0.333 |
| d | $np^2\,{}^3P$ | 0.301 |
| e | $np^2\,{}^1D$ | 0.29 |
| f | $np^2\,{}^1S$ | 0.342 |

## 2. Some experimental observations in gas phase atomic collisions and in particle surface interactions.

As mentioned above autodetaching states were first observed in electron scattering studies and have later been studied in photodetachment measurements and in heavy particle collisions. Recent work on ion and metastable atom scattering on surfaces has shown that these resonant states can be important intermediates in particle surface interactions and can be observed in electron spectroscopy studies. An extensive discussion of the observation of resonances in electron scattering and photodetachment can be found in the review by Buckman and Clark. In this section we give some examples of the observation of negative ion resonances in ion-atom collisions in order to illustrate the capabilities of this technique of negative ion spectroscopy and its usefulness in identifying decay channels. Some problems in early studies are pointed out. We then give some examples of electron spectroscopy studies of ion/atom surface collisions where these resonances were observed and discuss the role these can play.

### *2.1 Negative ion-atom/molecule collisions.*

General aspects of these collisions are extensively discussed in chapter 15. Autodetaching states of negative ions were first observed in an elegant series of electron spectroscopy studies (see e.g. the review by Risley 1979) of collisions of $H^-$, $C^-$, $O^-$ and halogen negative ion collisions with inert gasses performed by Bydin (1967), Risley et al (1974) and Edwards and Cunningham (1973,1974). More recently these were studied by the Paris/Orsay (Penent et al (1987,1988), Esaulov 1990), and Aarhus ( Andersen (1991), Andersen et al (1989), Dahl et al (1993), Poulsen et al (1990)) groups.

From the point of view of negative ion spectroscopy collisional excitation presents several attractive features amongst which are (i) the possibility of selective excitation of states (see chapter 15 for a discussion), (ii) the possibility of exciting states that can not be excited by photons because of selection rules and (iii) the possibility of studying very low energy electrons by using Doppler or kinematic shift of the electron energies in the laboratory frame of reference. Last but not least, these studies have given important indications concerning decay channels. Some of these features are illustrated in the following.

In an ion atom collision the excited negative ion is scattered with some (high) energy ($E_{ion}$ and velocity $v_{ion}$). In the laboratory frame of reference the energy of electrons ($E_0$ and velocity $v_0$) resulting from of autodetachment is therefore (Doppler) shifted to higher (or lower) energies (see e.g. Risley et al (1974)). For forward observation (i.e. at "zero" observation angle the electron energy can be strongly shifted to high energies ($m(v_e+v_{ion})2/2$). This shift allows one to observe emission of very low energy electrons, which are otherwise very hard to detect (for a detailed discussion see Penent et al 1991). A nice example of this is provided by recent studies of the decay of the $H^{-*}$ (2s2p) $^1P°$ shape resonance into H(n=2), leading to the emission of electrons of about 18meV( Andersen et al 1986, Duncan and Menendez 1989 and Penent et al 1991). Fig.1 shows results of measurements of Penent et al for 6keV $H^-$He collisions for forward (along the scattered H beam) angles. The bump at about 3.5eV is due to the shape resonance and corresponds to energies of about 20meV in the $H^-$ reference frame. As the observation angle decreases one can see the appearance of a double structure, clearly visible at 4°, due to electron emission into forward (high energy) and backward (low energy) directions. At very small angles one can observe a third, central peak (a cusp ) due to electrons of energies lower than 1meV. A detailed discussion of these features may be found in the above cited papers and the more recent work of Vikor et al (1996). This example nicely illustrates the utility of this kinematic shifting, which is frequently used in collision experiments. Similar experiments were recently reported for the $B^-$ case by Lee et al (1995). Note finally that from the point of view of state identification this effect allows one to distinguish between projectile and target states.

Let us now consider the question of decay channels. Fig 2a and b, show some electron spectra produced in $F^-$ and F collisions with He and Ne at low energies (Boumsellek et al 1990). The two lowest energy peaks in the $F^-$He spectrum (structures A and B in fig 2a,b) are due to the decay of the $F^-$ autodetaching state into excited states of F (Grouard et al 1986) as discussed above. All the peaks are forward shifted in energy from the centre of mass positions ( at about 0.1 and 0.25eV for A and B). Other structures are due to autoionising states of F . This attribution is confirmed by a comparative study of $F^-$ and F atom collisions (Boumsellek and Esaulov 1990) (fig.2b) where as may be seen the A and B peaks are not observed for the neutral projectile and are therefore related to $F^-$. This was also confirmed by photon-electron coincidence measurements (Poulsen et al 1990). Similar characteristics of decay of autoionising states of other halogen negative ions have been observed by Penent et al (1988). Note that these experiments show that it is important to perform studies of both negative ion and parent atom collisions in order to reach unambiguous identifications. Thus some confusion existed in the early studies of halogen negative ion collisions where some of the peaks in the electron spectra were incorrectly assigned to resonances, but were in fact due to decay of autoionising states (Penent et al 1988, Poulsen et al 1990, Andersen et al 1989). Corrected assignments are given in the tables below. We believe that in a number of cases this could have been due to the presence of a spurious neutral component in the negative ion beam (see also Penent et al 1988).

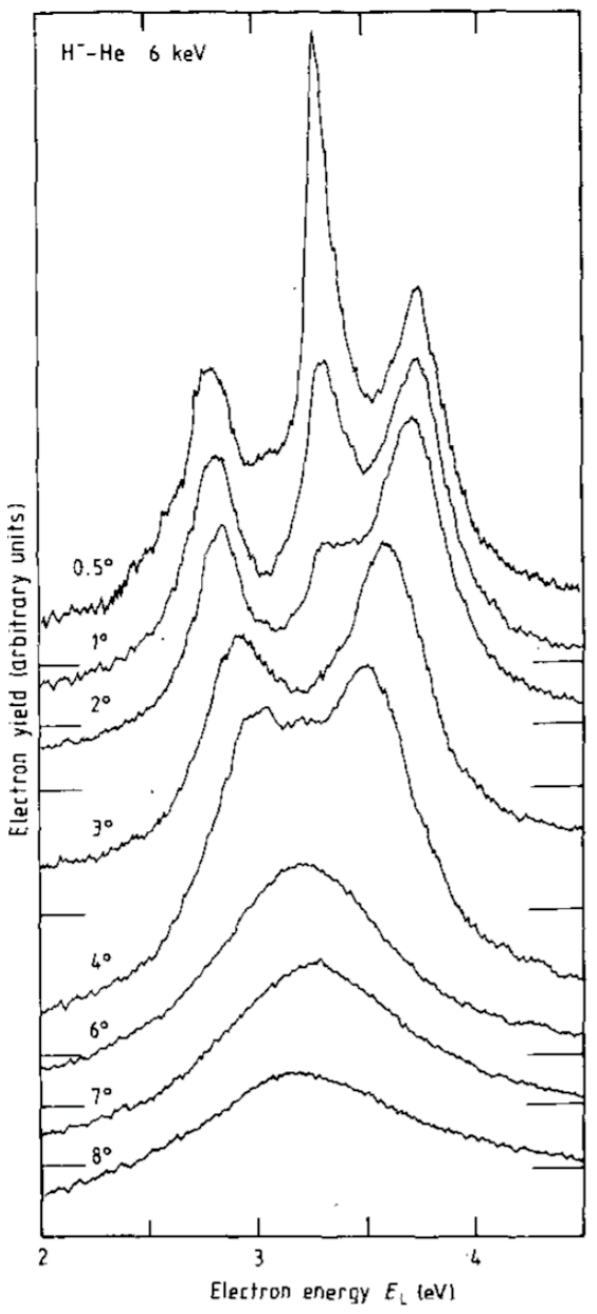

Fig.1. Electron energy spectra produced in $H^-He$ collisions for several observation angles with respect to the beam direction (from Penent et al 1991).

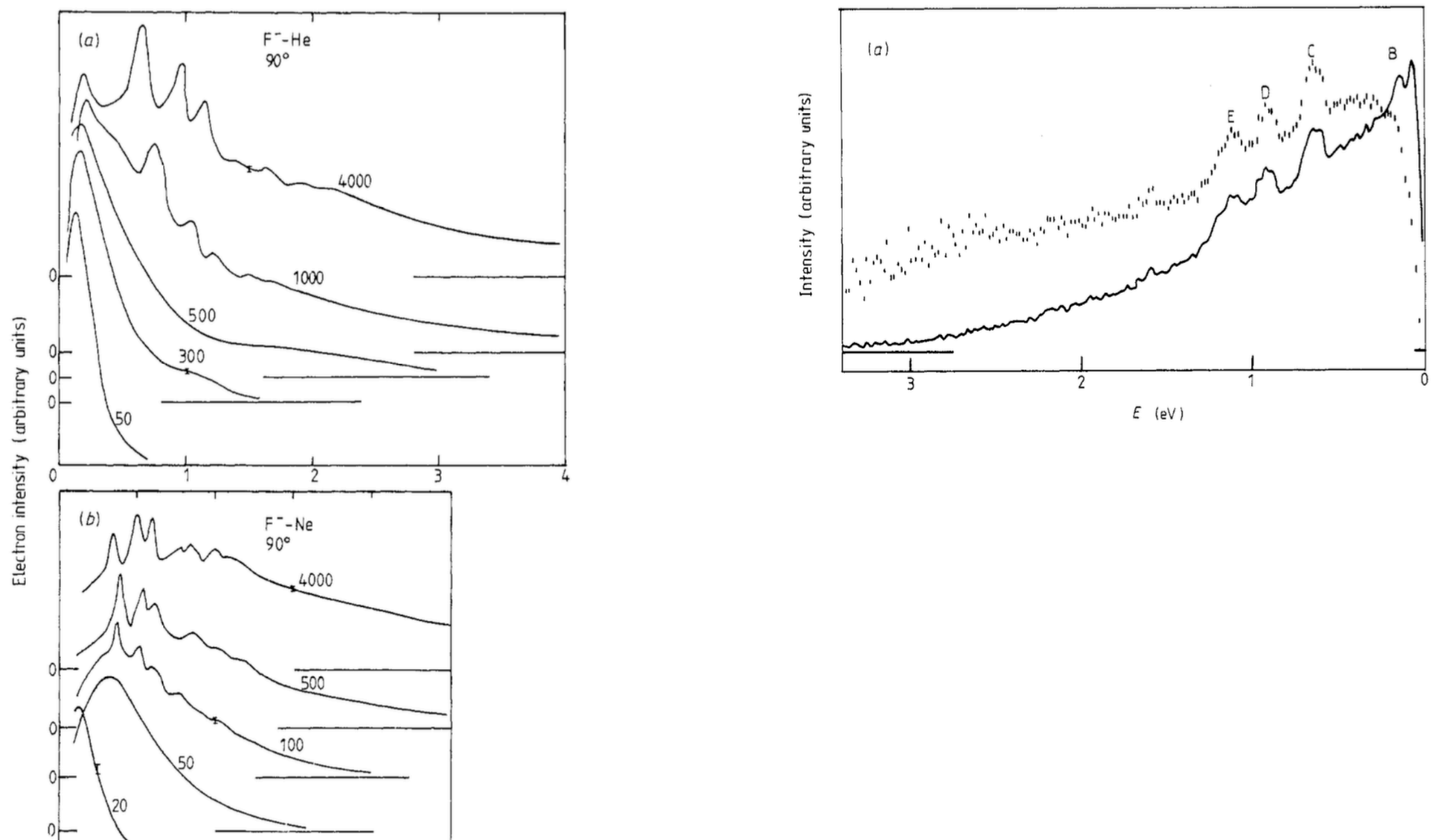

Fig.2. Detached electron energy spectra produced in $F^-$ collisions with He an Ne (left panel) and $F^-$ (solid line) and F (dashed line) collisions with He (right panel). Peaks A and B are due to $F^{-*}$ states. (from Grouard et al 1986 and Boumsellek and Esaulov 1990).

Another example is provided by the studies of $O^-$ (Penent et al 1987, Dahl et al (1993)) and $S^-$ collisions (Esaulov 1990), where the $O^{-*}$ $2p^3(^2D°)3s^2$ ($S^{-*}$ $3p^3(^2D°)4s^2$) resonance can decay into both the O $2p^4$ $^3P$ and O $2p^4$ $^1D$ states. Recent calculations of Dahl et al for $O^-$ give a 1.3 branching ratio for this decay favouring the $^1D$ channel. Fig 3. shows the electron spectra produced in $O^-$ and $S^-$ He collisions. The high energy peaks are due to decay into the ground state. A discrepancy in earlier work by Edwards and Cunningham (1973) should be pointed out here. Edwards and Cunningham reported main structures in their electron spectra at 10.11eV and 12.12 eV. They attributed the 10.11 eV peak to the decay of the $2s2p^6\ ^2S$ state into the $2p^4\ ^1D$ excited state of O. In the more recent work of Penent et al (1987, this data was not published in that paper) these dominant peaks are located at energies of 10.15±0.02 eV and 12.12 ±0.02 eV with an energy difference of 1.95 eV, which is in good agreement with the $^3P$ - $^1D$ splitting for the ground state configurations of O. This was pointed out by Dahl et al (1993) who present a very nice and thorough experimental and theoretical study and give calculated decay widths of $6.3*10^{-4}$ and $4.7*10^{-4}$eV for the $^1D$ and $^3P$ channels respectively.

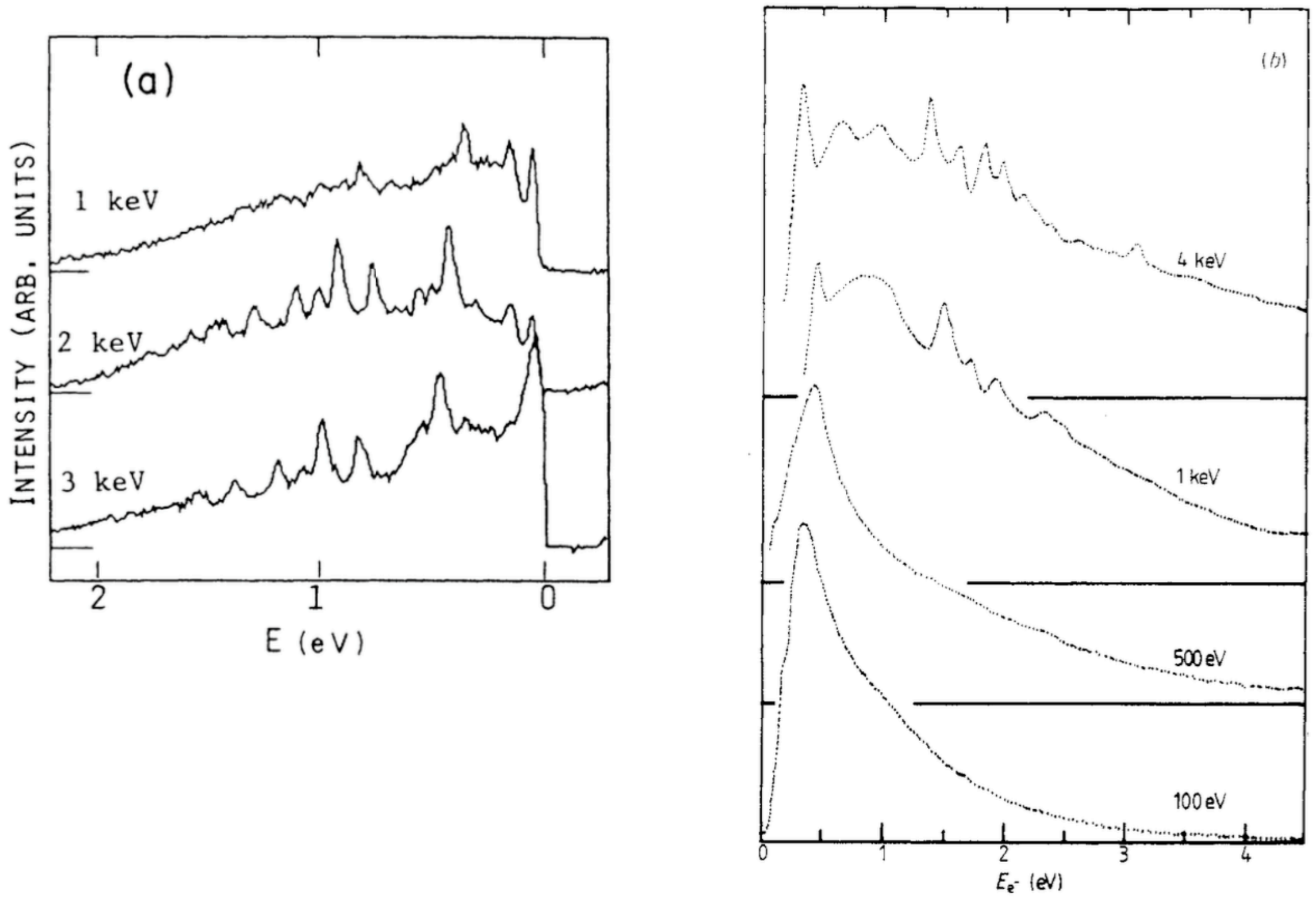

Fig.3. Electron energy spectra produced in (a) $S^-$ (Esaulov 1990) and (b) $O^-$ (Penent et al 1988) collisions with He. Energies are in the laboratory frame of reference.

Additional work on the study of excitation of resonances in negative ion collisions and calculations of branching ratios would be welcome.

## *2.2 Autodetaching states at surfaces.*

### *2.2.1 He $^-$* and other rare gases.*

In recent years it has been suggested that transient negative ion formation could play an important role in modifying the electronic structure of particles approaching the surface. We consider here the conversion of He* ($2^1S$) to He* ($2^3S$). One of the ways in which this can happen is via formation of autodetaching states of negative ions. Indeed some recent experimental MDS (metastable He deexcitation spectroscopy, a common tool in surface analysis) studies of deexcitation of He* ($2^1S$) in scattering on alkali covered surfaces has revealed the exciting evidence of involvement of the autodetaching $He^-$ ($1s2s^2\ ^2S$) state. The first report on this came from Hemmen and Conrad (1991), who were performing a study of a K covered Pt surface and observed a high energy peak in their spectrum which they assigned to the decay of $He^-$ ($1s2s^2\ ^2S$) formed near the surface. Similar observations were reported by Kempter and coworkers (Brenten et al 1992), who were studying He* deexcitation and neutralisation of $He^+$ and $He^{++}$ on alkali covered W(110). Fig. 4 shows the electron spectrum reported by Hemmen and Conrad (1991) case as a function of the K coverage. As may be seen as the coverage increases (with a *concomitant lowering of the surface workfunction*) a sharp structure appears in the high energy part of the spectrum, which is attributed to $He^-$*. Note that the energy calibration in the surface experiments is somewhat different from the "normal" practice in gas phase collisions and this is outlined in section 3.1.3

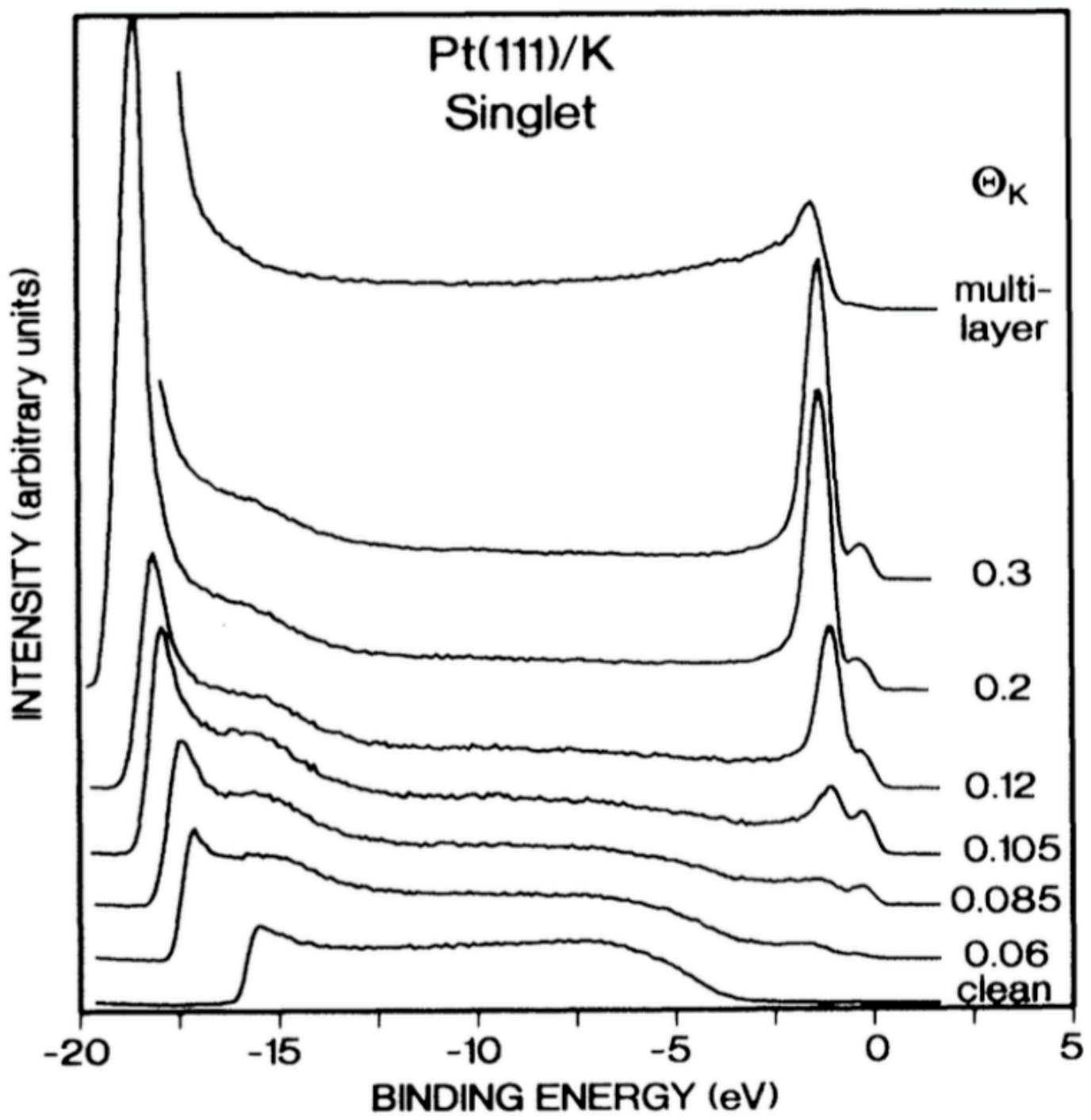

Fig.4. Electron energy spectra produced in He* interaction with a K covered Pt surface. (from Hemmen and Conrad 1991).

As initially discussed by Hemmen and Conrad (1991), this resonance is formed as a result of electron capture from the surface (general aspects of negative ion formation by electron capture are discussed extensively in chapters 20 and 21). The mechanism of capture and singlet to triplet rearrangement is illustrated in fig.5a, which shows a schematic representation of the binding energy of $He^{-}$* as a function of atom surface distance, assuming an image potential shift (see chapter 20, 21). The two curves in the figure correspond to the binding of the extra electron to He* ($2^1S$) (1.25 eV) and He* ($2^3S$) (0.45 eV). For a low workfunction surface the anion level will shift below the Fermi level at very large atom surface distances ($Z_S$) for the case of the He* ($2^1S$) parent, so that electron capture from the occupied levels of the surface can occur as He* ($2^1S$) approaches the surface, leading to $He^{-}$* formation. Autodetachment to the ground state, leads to the high energy electrons observed in fig.4a. In the case of $He^+$, the first step is neutralisation. This can occur by Auger neutralisation giving ground state He or resonantly giving He*(2s). He* can then capture an electron.

To understand the question of rearrangement, it is important to note now, that while the $He^{-}$* level related to He* ($2^1S$) lies *below* the Fermi level, *the $He^{-}$* level related to He* ($2^3S$) lies above it.*, and only shifts below the Fermi level much closer to the surface ($Z_t$). Therefore at intermediate distances, as $He^{-}$* continues its path towards the surface, electron loss to the conduction band can occur, leading to He* ($2^3S$).

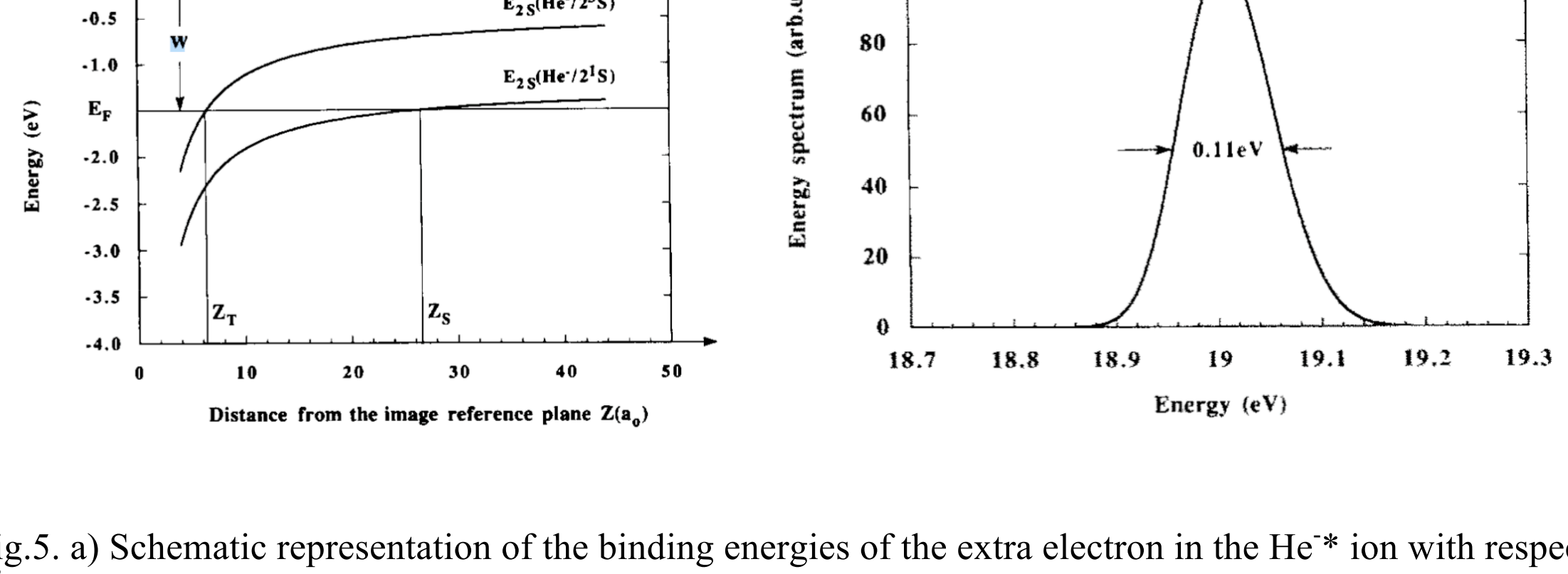

Fig.5. a) Schematic representation of the binding energies of the extra electron in the $He^{-}$* ion with respect to the He $2^1S$ and $2^3S$ levels as a function of atom surface distance.(from Borisov et al 1993).
b) Energy spectrum of electrons emitted during the decay of $He^{-}$* near the surface into the He ground state channel.

This process has been studied theoretically by Borisov et al (1993), who calculated the total and partial widths of the channels involved. Borisov et al show that the rearrangement process can indeed take place. In the context of negative ion spectroscopy, it is important to know what sort of electron spectrum would be obtained. Indeed since the levels are shifted near the surface the peak in the electron spectrum may not appear at the canonical 19.366 eV energy (see tables below). Borisov et al also calculated the population of the $He^{-}$* level near the surface and the shape of the electron spectrum that would be obtained. Fig.5b illustrates this shape for an 0.5eV collision energy. As may be seen the peak is shifted to lower energies and broadened, since decay can occur over a range of energies. It should be noted that if we were considering the movement of an excited atom *away* from the surface, a *different* type of spectrum could be obtained and in case of rapid movement the position of the peak could end up by being close to the "free atom" position. These features have been observed in case of production of autoionising states of Ne in the scattering of Ne ions on surfaces and a discussion of this may be found in the papers of Guillemot et al (1995) and Esaulov (1995).

Following the first observations for $He^{-}$* experiments with other rare gas ions (Ne, Ar) were performed by Muller et al (1993b), who observed similar structures in their electron spectra that were related to $Ne^{-}$*($3s^2$) and $Ar^{-}$* ($4s^2$).

*2.2.2. Reactive species : H, C, N and the case of metastable $N^{-}$.*

Experiments with reactive ions H, C and N have been performed by Muller et al (1993,1994,1996). Fig. 6a,b and c, shows some electron spectra for $H^+$, $C^+$ and $N^+$ scattering, where structures in the electron spectra are attributed to $H^{-}$* $C^{-}$* and $N^{-}$*. In case of $H^-$ the sharp peak in the 10 eV region is attributed to $H^{-}$* ($2s^2$). As for $He^+$ the first step in forming $H^{-}$* is the resonant neutralisation into H(n=2).

In case of $C^+$ neutralisation the electron spectrum displays three prominent structures, which appear progressively in the spectrum as the coverage of the alkali increases. The energy positions are 7.7eV, 6 eV and 4.6 eV. These are tentatively attributed to decay of the $2p3s^2$ $C^{-}$* into *different ground state levels of* C° ($2p^2$ $^3P$, $^1D$ and $^1S$). To this author's knowledge this seems to be the only observation of this $C^{-}$* resonance. Note that the calculated energy for decay into the ground state is 7.07 eV (see below), while the experimental value is obviously shifted to a too high energy.

It could be tempting to deduce branching ratios for decay into the different excited states from this spectrum. However the decay rates could be differently affected by the vicinity of the surface.

In case of $N^+$ one also observes several structures. The prominent series of peaks that appear at the higher coverages, i.e. for low workfunction values are attributed to decay of $N^{-}$*( $2p^2(^3P)3s^2$ into the *different ground state levels* of N ($2s^22p^3$ ($^4S$, $^2D$ and $^2P$).

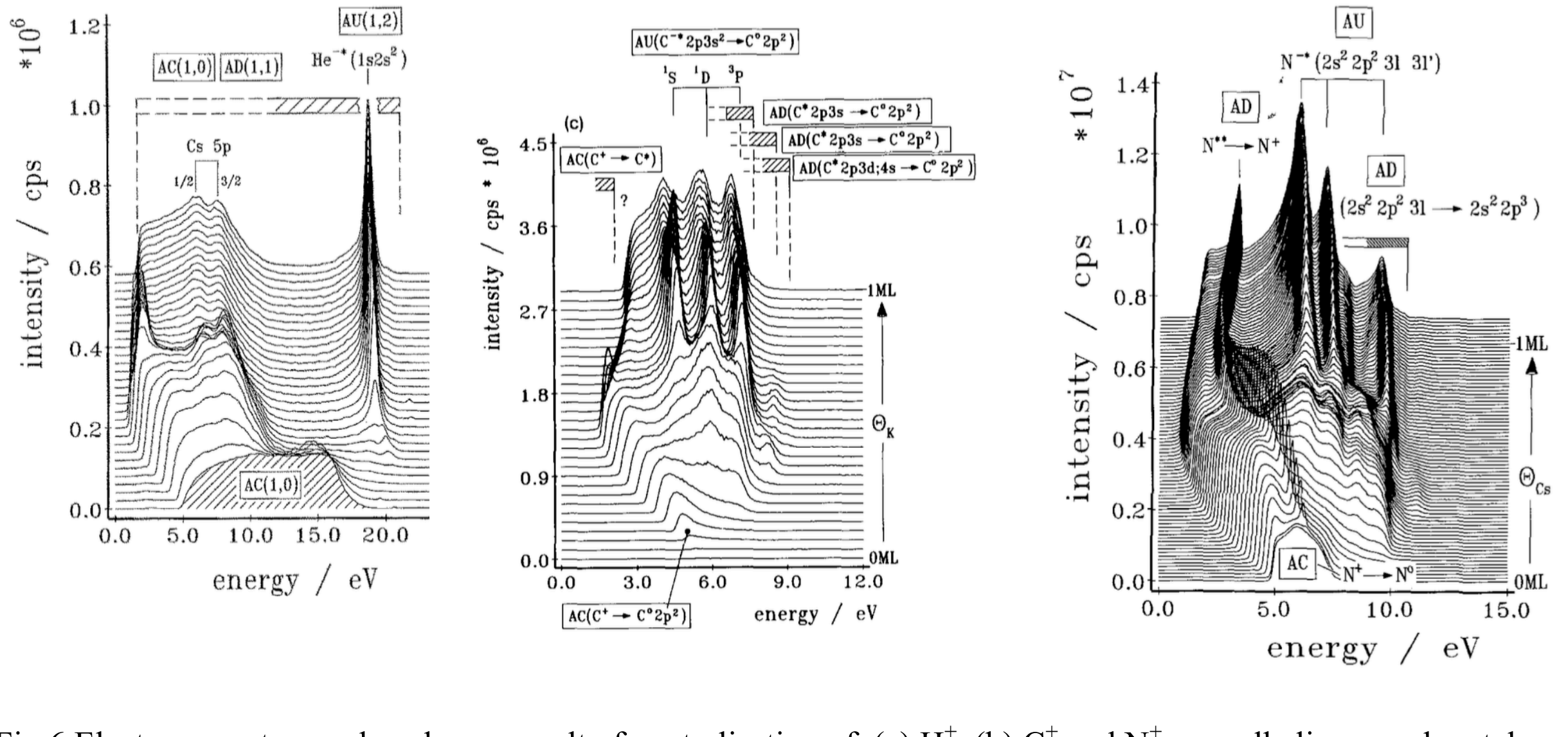

Fig.6 Electron spectra produced as a result of neutralisation of (a) $H^+$, (b) $C^+$ and $N^+$ near alkali covered metal surfaces as a function of alkali exposure. (from Muller et al 1993, 1994).

The other very interesting feature in the spectrum is the sharp peak at an energy of 1.4eV, which appears *as soon as* some alkali is present on the surface, i.e. well before the workfunction decreases. This peak was attributed to decay of the $N^{-}$* ($2p^4$ $^1D$) state, which lies above the ground state of N (see the schematic diagram in fig.7). Note that this is bound with respect to the other states of the ground configuration of N and can hence be formed by resonant capture involving e.g. N($^2D$), which is itself again formed as a result of neutralisation of $N^+$. Note here that neutralisation of $N^+$ can lead to the different ground state levels ( $^4S$, $^2D$ and $^2P$ ) and into excited states (like N*(3s)).

The curious feature here is that this peak appears, when the workfunction is quite large, i.e. in conditions when for a *high workfunction surface* one would certainly not expect an efficient production of such a weakly bound species (see chapter 21). The answer to this problem was provided in a recent theoretical investigation by Bahrim et al (1997), who performed a detailed calculation of the position and width of $N^{-*}$ ($2p^4\ ^1D$) in front of the alkali adsorbate at low coverages. A discussion of this is beyond the scope of this section, but the main result should be pointed out since it is of fundamental general importance in surface chemistry. This is that the widths and positions of this state in front of this electropositive adsorbate are *totally different* from what they would be in front of a jellium like metal. This occurs because of a *locally* determined interaction with the adsorbate and does not rely on the lowering of a globally determined workfunction (non-local effect). This example provides a vivid demonstration of adsorbate induced modifications in surface reactivity, which have been the object of much experimental (see e.g. the review by Rodriguez and Goodman 1992) and theoretical work (Norskov et al (1984), Lang et al (1985), Feibelman and Hamman (1984), Nordlander and Lang (1991)) in the past. Studies of negative ion formation thus provide an excellent opportunity to study these effects and are the object of intensive work at the time of writing.

The above mentioned experiments on $N^-$ stimulated a recent theoretical investigation by Cowan et al (1997), who calculated the positions of the $N^-$ states as indicated in the tables below.

A number of other experiments involving molecular ions have been performed, where the formation of autoionising states of both atomic and molecular species is observed. Müller et al (1994) have observed production of both $N^{-*}$ ($2p^4\ ^1D$) and attribute some structures in their spectra to $N_2^{-*}$ ($^2\Pi g$) and $N_2^{-*}$ ($E^2\Sigma g^+$).

Additional work in this field could be a useful means of obtaining additional spectroscopic information on excited negative ion states. If in fast ion scattering these states were populated during the outgoing path, as the atom moves away from the surface, then the energy position of the electron peaks would not be affected by energy level shifts near the surface and could be very close to the free atom energy.

### *2.2.3 The energy calibration at surfaces.*

The object of this paragraph is to point out a specific feature of observing decay of excited states of atoms near surfaces, whose workfunctions vary as in the case of alkali adsorption discussed above. One considers that the Fermi levels of the spectrometer and target surface are tied together. The energy ($E_0$) of the electron emitted by an *excited atom* near a surface is related to the vacuum level of the target. Therefore the energy measured ($E_m$) by the spectrometer is

$$E_m = E_0 - \delta\Phi$$

It follows that if the workfunction of the target decreases, $E_m$ decreases. This explains why the positions of the peaks in the electron spectra change.

Note furthermore that it is quite common practice to polarise a surface, in such a manner that zero energy electrons appear at some positive energy. The low energy cutoff of the spectrum gives the workfunction variation. The energy calibration normally requires the knowledge of the spectrometer and target workfunctions. This is not a very simple matter and in such experiments is rarely determined to better than some 50 to 100meV, i.e. the ensuing calibration is worse than the calibrations obtained in gas phase scattering with reference to some well known resonance like the inert gas resonances.

## 3. Tabulated data on Atomic negative ion resonances.

A summary of some known excited states of atomic negative ions is presented in the following tables, for states below the first ionisation limit. With few exceptions, the identifications of the negative ion resonances and their energies are as given in the review of Buckmann and Clark. Details about the identification, the precision with which energies are determined and the widths of the states may be found in their review. Some additional data is given here with references.

The energies of the states are given (in eV) with respect to the ground state of the parent atom.

**Table 1**

***Hydrogen***

| State | Energy (eV) |
|---|---|
| $2s^2\ ^1S$ | 9.557 |
| $2s2p\ ^3P°$ | 9.735 |
| $2p2\ ^1De$ | 10.115 |
| $2s2p\ ^1P°$ | 10.172 |
| | 10.226 |
| $^1S$ | 11.722 |
| $^1D$ | 11.807 |
| $3s3p\ ^1P°$ | 11.9 |
| $^3F°$ | 11.925 |
| $^3D$ | 11.997 |
| $^1S$ | 12.029 |
| $^3P$ | 12.036 |
| $^1D$ | 12.049 |
| $3s3p\ ^1P$ | 12.082 |

***Helium***

| State | Energy (eV) |
|---|---|
| $2s^2\ ^2S$ | 19.366 |
| $2s2p\ ^2P$ | 20.4 |
| $2p^2\ ^4P^e$ | 20.963 |
| $2p^2\ ^2D$ | 21.0 |
| $3s^2\ ^2S$ | 22.44 |
| $3s3p\ ^2P$ | 22.6 |
| $3s3d\ ^2D$ | 22.66 |
| $3s\ p\ ^2P$ | 22.74 |
| $3s\ s\ ^2S$ | 22.88 |
| $3p^2\ ^2D$ | 22.93 |
| $^2F$ | 22.99 |
| $^2G$ | 23.05 |

***Sodium***

| State | Energy (eV) |
|---|---|
| $3s3p\ ^3P°$ | 0.08 |
| $3p4s\ ^1P°+3p^2\ ^1D$ | 2.1 |
| $3p3d\ ^3F°$ | 2.57 |
| $4s2\ ^1S$ | 3.2 |
| $4s3d\ ^1D$ | 3.53 |

***Lithium***

The Li shape resonance data is from Lee et al (1996).

| State | Energy (eV) |
|---|---|
| $2s2p\ ^3P°$ | 0.05 |
| 2p3s 1P°+ $2p^2$ $^1D$ | 1.85 |
| $^1S$, $^3P$, $^3S$ | 3.11-3.34 |
| $^1D$, $^1S$, $^3P$, $^1P$ | 3.7-3.78 |

***Potassium***

| State | Energy (eV) |
|---|---|
| 4s4p $^3P°$ | 0.01 (?) |
| 4p5s $^1P°$ + $4p^2$ $^1S$ | 1.61 |
| 4p3d $^3F$ | 1.86 |
| | 2.4 |
| 5s4d $^1D$ | 2.6-2.63 |
| P or F | 2.68 |

***Rubidium***

| State | Energy (eV) |
|---|---|
| 5s5p $^3P°$ | <0.05 |
| 5p6s $^1P°$ | 1.559 |
| 5p6s $^1P°$ | 1.588 |
| 5p2 $^1S$- $5p^2$ $^1D$ | 1.58 |
| 5p4d $^3F°$ | 1.74 |
| 6s5d $^1D$ | 2.4 |
| $7s^2$ $^1S$ | 2.45-2.5 |

***Cesium***

| State | Energy (eV) |
|---|---|
| 6s6p $^3P°$ | 0.15 |
| 6p7s $^1P°$ + 6p5d 1P° | 1.38 |
| | 1.45 |
| $6p^2$ $^1S$, $^1D$ | 1.39 |
| | 1.45 |
| 6p5d $^3F°$ | 1.49 |
| 6p5d $^3D°$ | 1.8 |
| 6s5d $^1D$ | 2.2-2.3 |

***Neon***

| State | Energy (eV) |
|---|---|
| $2p^5$ ($P°_{3/2}$) $3s^2$ ($^1S$) | 16.11 |
| $2p^5$ ($P°_{1/2}$) $3s^2$ ($^1S$) | 16.208 |
| 3s3p (3P°) | 16.906 |
| 3s3p 1P° | 18.35 |
| 3p2 1D | 18.464 |
| $2p^5$ ($P°_{3/2}$) $3p^2$ ($^1S$) | 18.58 |
| $3p^2$ ($^1D$) | 18.626 |
| $2p^5$ ($P°_{1/2}$) $3p^2$ ($^1S$) | - |
| 3p2 1D | 18.672 |
| 3pl+es | 18.965 |
| $2p^5$ ($P°_{3/2}$) $4s^2$ ($^1S$) | 19.498 |
| $2p^5$ ($P°_{1/2}$) $4s^2$ ($^1S$) | 19.598 |
| $2p^5$ ($P°_{3/2}$) $4d^2$ ($^1S$) | 19.686 |
| $2p^5$ ($P°_{1/2}$) $4d^2$ ($^1S$) | 1.778 |
| $2p^5$ ($P°_{3/2}$) $4p^2$ ($^1S$) | 20.054 |
| $2p^5$ ($P°_{1/2}$) $4p^2$ ($^1S$) | 20.15 |

***Argon***

The values labelled with an * are "benchmark" energies determined with 1meV accuracy by Hammond (1996).

| State | Energy (eV) |
|---|---|
| $3p^5(^2P°_{3/2})\ 4s^2\ (^1S)$ | 11.1030* |
| $3p^5(^2P°_{1/2})\ 4s^2\ (^1S)$ | 11.2763* |
| $3p^5(^2P°_{1/2})\ 4s4p\ (^3P°)$ | 11.631 |
| | 11.675 |
| | 11.785 |
| | 11.845 |
| $3p^5(^2P°_{3/2,\ 1/2})\ 4s4p\ (^1P°)$ | 12.7 |
| $3p^5(^2P°_{3/2,\ 1/2})\ 4p^2\ (^1D)$ | 12.925 |
| | 12.942 |
| | 12.99 |
| $3p^5(^2P°_{3/2})\ 4p^2\ (^1S)$ | 13.055 |
| $3p^5(^2P°_{3/2,\ 1/2})\ 4p^2\ (^1D)$ | 13.162 |
| | 13.19 |
| $3p^5(^2P°_{1/2})\ 4p^2\ (^1S)$ | 13.216 |
| $3p^5(^2P°_{3/2,\ 1/2})\ 4p^2\ (^1D)$ | 13.282 |
| $3p^5(^2P°_{1/2})$ 4pl° +es | 13.479 |
| $3p^5(^2P°_{3/2})\ 3d^2\ (^1S)$ | 13.864 |
| | 13.907 |
| $3p^5(^2P°_{3/2})\ 4p^2\ (^1S)$ | 14.006 |
| $3p^5(^2P°_{3/2})\ 5s^2\ (^1S)$ | 14.052 |
| $3p^5(^2P°_{1/2})\ 5s^2\ (^1S)$ | 14.214 |
| | 14.434 |
| | 14.478 |
| $3p^5(^2P°_{3/2})\ p^2\ (^1S)$ | 14.53 |
| $3p^5(^2P°_{3/2,\ 1/2})\ 5pl^2$ +es | 14.556 |
| | 14.634 |
| $3p^5(^2P°_{1/2})\ 5p^2\ (^1S)$ | 14.711 |
| $3p^5(^2P°_{\ 1/2})\ 5pl_0°$ + es | 14.735 |
| $6l^2\ (^1S)$ | 14.811 |

***Krypton***

| State | Energy (eV) |
|---|---|
| $4p^5(^2P°_{3/2})\ 5s^2\ (^1S)$ | 9.484 |
| $4p^5(^2P°_{3/2})\ 5s5p\ (^3P°)$ | 10.039 |
| $4p^5(^2P°_{1/2})\ 5s^2\ (^1S)$ | 10.119 |
| $4p^5(^2P°_{1/2})\ 5s5p\ (^3P°)$ | 10.6 |
| $4p^5(^2P°_{3/2})\ 5s5p\ (^1P°)$ | 11.12 |
| $4p^5(^2P°_{3/2})\ 5p^2\ (^1D)$ | 11.286 |
| $4p^5(^2P°_{3/2})\ 5p^2\ (^1D)$? | 11.318 |
| $4p^5(^2P°_{3/2})\ 5p^2\ (^1S)$ | 11.4 |
| $4p^5(^2P°_{3/2})\ 5p^2\ (^1D)$ | 11.49 |
| $4p^5(^2P°_{3/2})$ 5pl +es | 11.653 |
| $4p^5(^2P°_{1/2})\ 5s5p\ (^1P°)$ | 11.77 |
| $4p^5(^2P°_{3/2})\ 5p^2\ (^1D)$ | 11.996 |
| $4p^5(^2P°_{1/2})\ 5p^2\ (^1S)$ | 12.036 |
| $4p^5(^2P°_{1/2})\ 5p^2\ (^1D)$ | 12.138 |
| | 12.191 |
| $4p^5(^2P°_{1/2})$ 5pl +es | 12.262 |
| $4p^5(^2P°_{3/2})\ 6s^2\ (^1S)$ | 12.378 |
| $4p^5(^2P°_{3/2})\ 5p^2\ (^1S)$ | 12.76 |
| | 12.875 |
| $4p^5(^2P°_{1/2})\ 6s^2\ (^1S)$ | 13.016 |
| | 13.067 |
| $4p^5(^2P°_{3/2})\ 7l^2\ (^1S)$ | 13.221 |
| | 13.291 |
| | 13.379 |
| $4p^5(^2P°_{1/2})\ 5p^2\ (^1S)$?? | 13.43 |
| | 13.477 |
| $4p^5(^2P°_{3/2})\ 8l^2\ (^1S)$ | 13.528 |
| $4p^5(^2P°_{3/2})\ 9l^2\ (^1S)$ | 13.598 |
| $4p^5(^2P°_{3/2})\ 10l^2\ (^1S)$ | 13.714 |
| $4p^5(^2P°_{3/2})\ 11l^2\ (^1S)$ | 13.789 |
| $4p^5(^2P°_{1/2})\ 7l^2\ (^1S)$ | 13.891 |
| $4p^5(^2P°_{1/2})\ 8l^2\ (^1S)$ | 14.199 |
| $4p^5(^2P°_{1/2})\ 9l^2\ (^1S)$ | 14.273 |
| $4p^5(^2P°_{1/2})\ 10l^2\ (^1S)$ | 14.375 |
| $4p^5(^2P°_{1/2})\ 11l^2\ (^1S)$ | 14.441 |

***Xenon***

| State | Energy (eV) |
|---|---|
| $5p^5(^2P°_{3/2})\ 6s^2\ (^1S)$ | 7.90 |
| $5p^5(^2P°_{3/2})\ 6s6p\ (^3P°)$ | 8.338 |
| $5p^5(^2P°_{3/2})\ 6s6p\ (^3P°)$ | 8.43 |
| $5p^5(^2P°_{3/2})\ 6s6p\ (^1P°)$ | 9.08 |
| $5p^5(^2P°_{1/2})\ 6s^2\ (^1S)$ | 9.18 |
| $5p^5(^2P°_{3/2})\ 6s6p\ (^1P°)$ | 9.36 |
| $5p^5(^2P°_{3/2})\ 6p^2\ (^1D)$ | 9.505 |
| $5p^5(^2P°_{3/2})\ 6p^2\ (^1D)$ | 9.551 |
| $5p^5(^2P°_{3/2})\ 6p^2\ (^1S)$ | 9.623 |
| | 9.644 |
| | 9.686 |
| | 9.743 |
| | 9.831 |
| | 9.896 |
| | 10.48 |
| $5p^5(^2P°_{1/2})\ 6s6p\ (^1P°)$ | 10.71 |
| $5p^5(^2P°_{3/2})\ 6s6p\ (^1P°)$ | 10.858 |
| | 10.901 |
| $5p^5(^2P°_{1/2})\ 6p^2\ (^1S)$ | 10.957 |

***Fluorine***

The $2p^4(^1D)\ 3s^2$ state has been observed in $F^-$ collisions. Other data are calculated values.

| State | Energy (eV) |
|---|---|
| $2p^4(^3P)\ 3s^2$ | 12.29 |
| $2p^4(^1D)\ 3s^2$ | 14.85 |
| $2p^4(^1S)\ 3s^2$ | 17.69 |

***Chlorine***

Observation of most of the structures in the electron spectra of Cunningham and Edwards (1973) is not confirmed by later measurements of Penent et al (1988) and Andersen et al (1989). A number of these were due to autoionising states of Chlorine. In the later works only the lowest $3p^4(^1D)\ 4s^2$ resonace is observed in collisions with inert gasses. The following table reflects this. We have left the $3p^4(^3P)\ 4s^2$ state since it was reported for collisions with $H_2$. The observation of the $3p^4(^3P)\ 4s^2$ state is unlikely in low energy $Cl^-$ rare gas collisions for reasons outlined in chapter 15.

| State | Energy (eV) |
|---|---|
| $3p^4(^3P)\ 4s^2$ | 8.53 ? |
| $3p^4(^1D)\ 4s^2$ | 9.97 |

***Bromine***

The energies of $Br^-$ resonances are taken from the more recent and accurate work of Penent et al (1988), than the data of Edwards and Cunningham (1974) used by Buckman and Clark. Penent et.al present a brief discussion of excited states and decay channels of autodetaching states of halogen negative ions produced in collisions with inert gasses.

| State | Energy (eV) |
|---|---|
| $4p^4(^3P_2)\ 5s^2$ | 7.50 |
| $4p^4(^3P_{1,0})\ 5s^2$ | 7.95 |
| $4p^4(^1D_2)\ 5s^2$ | 8.97 |

***Iodine***

The production of the $5p^4(^3P_2)\ 6s6p\ (^3P°)$ state in $I^-$ collisions with He, reported by Cunnigham and Edwards (1974) is not confirmed by Penent et al (1988).

| State | Energy (eV) |
|---|---|
| $5p^4(^3P_2)\ 6s^2$ | 6.41 |
| $5p^4(^3P_{1,0})\ 6s^2$ | 7.15 |
| $5p^4(^1D_2)\ 6s^2$ | 8.06 |

***Boron***

$B^-$ scattering by He and Ar was studied recently by Lee et al (1995). They report observation of several structures in the electron spectra, one of which was tentatively identified with the $2s^22p^2\ ^1D$ shape resonance. The energy position appears in agreement with theoretical estimates of Froese-Fischer (see Lee et al).

| State | Energy (eV) |
|---|---|
| $2s^22p^2\ ^1D$ | 0.104 |

***Carbon***

The $2p\ 3s^2\ (^2P°)$ state and its decay has been observed for collisions with surfaces (see §2 above). We do not include the energy positions of the corresponding peaks because of the likelihood of surface induced shifts as outlined above. Calculated values due to Matese[a] and Clark[b] (see Buckman and Clark 1994) are indicated

| State | Energy (eV) |
|---|---|
| $2s^2\ 2p^3\ ^2P°$ | 0.57 |
| | 0.46 |
| $2s2p^4\ ^4P$ | 6.1 |
| $2p\ 3s^2\ (^2P°)$ | 7.07 [a] |
| | 7.05 [b] |
| $2p(^2P°)\ 3s3p(^3P°)$ | 7.44 |

***Nitrogen***

Excited states of nitrogen $2p^4$ ($^1D$) and $2p^4$ ($^1S$) were recently discussed by Cowan *et al* (1997) in relation to experiments of $N^+$ scattering on surfaces on alkalated surfaces. *Calculated* energies of these states are indicated here. Values of earlier calculations may be found in the work of Cowan *et al.* $N^{-}$*( $2p^2(^3P)3s^2$ has been observed in $N^+$ surface scattering (see fig6.). Not included here for the same reason as for C.

| State | Energy (eV) |
|---|---|
| $2p^4$ ($^3P$) | 0.07 |
| | 0.065-0.09 |
| $2p^4$ ($^1D$) | 1.513 |
| $2p^4$ ($^1S$) | 2.903 |

***Oxygen.***

The only resonance excited in $O^-$ He collisions is the $2p^3(^2D°)3s^2$ one. Excitation of the $2s2p^6\ ^2S$ state is not confirmed (see §2.1 above). Other data are from Spence (1975)

| State | Energy (eV) |
|---|---|
| $2p^3(^4S°)\ 3s^2$ | 8.78 |
| $3s3p(^3P°)$ | 9.5 |
| $3p^2\ ^4P°$ | 10.73 |
| $3p^2\ ^2P°$ | 10.87 |
| $2p^3(^2D°)3s^2$ | 12.12 |
| $2p^3(^2D°)\ 3s3p(^3P°)$ | 12.55 |
| $2p^3(^2P°)\ 3s^2$ | 13.71 |
| $2p^3(^2D°)3p^2$ | 14.05 |
| $2p^3(^2P°)3s3p(^3P°)$ | 14.4 |
| $2p^3(^2D°)3p^2$ | 15.65 |

***Sulphur***

A preliminary electron spectroscopy study of autodetaching state formation in $S^-$ collisions with He (Esaulov 1990) showed a peak at an energy of $8.05 \pm 0.05$ eV attributable to the $4s^2$ resonance.

| State | Energy (eV) |
|---|---|
| $3p^3(^2D°)4s^2$ | $8.05 \pm 0.05$ |

***Zinc***

| State | Energy (eV) |
|---|---|
| $3d^{10}4s^24p\ ^2P°$ | 0.49 |
| $3d^{10}4s^24p\ ^2D$ | ≈2.5 |
| $3d^{10}4s4p^2$ | 4.36 |
| $3d^{10}4s5s^2$ | 6.0-6.4 |
| | 7.18 |
| $3d^{10}4s5p^2$ | 7.56 |

***Cadmium***

| State | Energy (eV) |
|---|---|
| $4d^{10}5s^25p\ ^2P°$ | 0.33 |
| $4d^{10}5s^25d\ ^2D$ | ≈2.0 |
| $4d^{10}5s5p^2$ | 4.16 |
| $4d^{10}5s5p^2$ | 5.42 |
| $4d^{10}5s6s^2$ | ≈5.9 |
| $4d^{10}5s6p^2$ | 6.82 |
| $4d^{10}5s6p^2$ | 7.24 |
| $4d^{10}5s6p^2$ | 7.38 |

***Mercury***

| State | Energy (eV) |
|---|---|
| $5d^{10}6s^26p\ ^2P°$ | 0.4-0.5 ?? |
| $5d^{10}6s6p^2\ ^4P_{1/2}$ | 4.55 |
| $5d^{10}6s6p^2\ ^4P_{3/2}$ | 4.702 |
| $5d^{10}6s6p^2\ ^4P_{5/2}$ | 4.94 |
| $5d^{10}6s6p^2\ ^2D_{3/2}$ | 5.2 |
| $5d^{10}6s6p^2\ ^2D_{5/2}$ | 5.54 |
| $5d^{10}6s6p^2\ ^2P_{1/2}$ | 6.3-6.8 |
| $5d^{10}6s6p^2\ ^2S_{1/2}$ | 6.702 |
| $5d^{10}6s7s^2\ ^2S_{1/2}$ | 7.5 |
| $5d^{10}6s6p^2\ ^2P_{3/2}$ | 7.6-8.1 |
| $5d^96s^2\ (^2D_{5/2})\ 6p^2$ | 8.367 |
| | 8.508 |
| $5d^{10}6s7p^2$ | 8.56 |
| | 8.65 |
| $5d^96s^2\ (^2D_{5/2})\ 6p^2$ | 8.854 |
| $5d^96s^2\ (^2D_{5/2})\ 6p^2$ | 9.439 |
| $5d^96s^2\ (^2D_{5/2})\ 6p^2$ | 9.593 |
| | 9.988 |
| $5d^96s^2\ (^2D_{5/2})\ 6p^2$ | 10.356 |
| $5d^{10}6s^26p^2$ | 10.612 |
| $5d^{10}6s^26p^2$ | 10.9 |
| $5d^{10}6s^26p^2$ | 11.05 |
| $5d^{10}6s^26p^2$ | 11.3-11.5 |
| $5d^{10}6s^26p^2$ | 11.788 |
| $5d^{10}6s^26p^2$ | 12.095 |
| $5d^96s^2\ (^2D_{5/2})\ 7p^2$ | 12.852 |
| $5d^96s^2\ (^2D_{5/2})\ 7p^2$ | 12.989 |
| $5d^96s^2\ (^2D_{5/2})\ 7p^2$ | 13.041 |
| $5d^96s^2\ (^2D_{5/2})\ 7p^2$ | 13.19 |
| $5d^96s^2\ (^2D_{5/2})\ 7p^2$ | 13.325 |
| $5d^96s^2\ (^2D_{5/2})\ 8s8p$ | 13.575 |
| $5d^96s^2\ (^2D_{5/2})\ 8s8p$ | 13.625 |
| | 13.718 |
| $5d^96s^2\ (^2D_{5/2})\ 8p^2$ | 13.881 |
| $5d^96s^2\ (^2D_{5/2})\ 8p^2$ | 13.944 |
| $5d^96s^2\ (^2D_{5/2})\ 9s9p$ | 14.034 |

**References.**

**Table 2.** Electron affinities of atomic negative ions. The numbers in parenthesis correspond to the errors in the last decimal fugures (e.g. $Fe^-$ 0.164 (35) stands for (0.164 ± 0.035)eV. Metastable states are denoted by the letter *m* (see also Table 2.). This negative ion data in this table is based on: V.Esaulov, *Ann Phys Fr* 11, 493-592, 1986. Some updated data is given according to other references at the end of the section. See also ref 1 for a general update.

| Z | Atom | Ionisation energy | Negative Ion | Electron affinity (eV) |
|---|---|---|---|---|
| 1 | H 1s $^2S_{1/2}$ | 13.5984 | $1s^2$ $^1S_0$ | 0.754209 (3) |
| 2 | He $1s^2$ $^1S_0$ | 24.5874 | | < 0 |
| | He 1s2s $2^3S$ | | 1s2s2p $^4P°$(m) | 0.078 (5) |
| 3 | Li 2s $^2S_{1/2}$ | 5.3917 | $2s^2$ $^1S_0$ | 0.6173(7) |
| 4 | Be $2s^2$ $^1S_0$ | 9.3227 | $2s^2$ 2p $^2P$ | <0 |
| | | | $2s^2$ 3s $^2S$ | < 0 |
| | Be 2s2p $^3P$ | | 2s $2p^2$ $^4P^e$(m) | 0.24 (10) |
| 5 | B 2p $^2P_{1/2}$ | 8.298 | $2p^2$ $^3P_0$ | 0.278(10) |
| 6 | C $2p^2$ $^3P_0$ | 11.2603 | $2p^3$ $^4S_{3/2}$ | 1.269 (3) |
| | | | $2p^3$ $^2D$ (m) | 0.033 |
| 7 | N $2p^3$ $^4S_{3/2}$ | 14.5341 | $2p^4$ $^3P$ | -0.07 (8) |
| | N $2p^3$ $^2D$ | | $2p^4$ $^1D$ (m) | 1.0 (3) |
| | N $2p^3$ $^2P$ | | $2p^4$ $^1S$ (m ?) | 0.9 (3) |
| 8 | O $2p^4$ $^3P_{3/2}$ | 13.6181 | $2p^5$ $^2P_{3/2}$ | 1.461 112(3)[2] |
| 9 | F $2p^5$ $^2P_{3/2}$ | 17.4228 | $2p^6$ $^1S_0$ | 3.401 1895 (25)[3,4] |
| 10 | Ne $2p^6$ $^1S_0$ | 21.5646 | | <0 |
| 11 | Na 3s $^2S_{1/2}$ | 5.1391 | $3s^2$ $^1S_0$ | 0.546 (5) |
| 12 | Mg $3s^2$ $^1S_0$ | 7.6462 | $3s^2$ 3p $^2P$ | < 0 |
| | Mg $3s^2$ $^1S_0$ | | $3s^2$ 4s $^2S$ | < 0 |
| | Mg 3s3p $^3P_0$ | | 3s $3p^2$ $^4P^e$ (m) | 0.32 (10) |
| 13 | Al 3p $^2P_{1/2}$ | 5.9858 | $3p^2$ $^3P_0$ | 0.442 (10) |
| | Al 3p $^2P_{1/2}$ | | $3p^2$ $^1D_2$ (m) | 0.109 (10) |
| 14 | Si $3p^2$ $^3P_0$ | 8.1517 | $3p^3$ $^4S_{3/2}$ | 1.389 5220(24)[4] |
| | Si $3p^2$ $^3P_0$ | | $3p^3$ $^2P_{3/2,5/2}$ (m) | 0.523 (5) |
| | Si $3p^2$ $^3P_0$ | | $3p^3$ $^2P_{1/2,3/2}$ (m) | 0.029 (5) |

| Z | Atom | Ionisation energy | Negative Ion | Electron affini |
|---|---|---|---|---|
| 15 | P $3p^3\ {}^4S_{3/2}$ | 10.4867 | $3p^4\ {}^3P_2$ | 0.7464 (4) |
| | P $3p^3\ {}^4S_{3/2}$ | | $3p^4\ {}^1D_2$ | ~0 |
| 16 | S $3p^4\ {}^3P_2$ | 10.36 | $3p^5\ {}^2P_{3/2}$ | 2.0772 (5) |
| 17 | Cl $3p^5\ {}^2P_{3/2}$ | 12.9676 | $3p^6\ {}^1S_0$ | 3.615 (4) |
| 18 | Ar $3p^6\ {}^1S_0$ | 15.7596 | | < 0 |
| | Ar $3p^5\ 4s\ {}^3P^0$ | | $3p^5\ 4s4p\ {}^4S$(m) | 0.135 |
| 19 | K $4s\ {}^2S_{1/2}$ | 4.3407 | $4s^2\ {}^1S_0$ | 0.50147 (10) |
| 20 | Ca $4s^2\ {}^1S_0$ | 6.1132 | $3d\ 4s^2\ {}^2D$ | < 0 |
| 21 | Sc $3d4s^2\ {}^2D_{3/2}$ | 6.5615 | $3d^2 4s^2\ {}^3F$ | < 0 |
| 22 | Ti $3d^2\ 4s^2\ {}^3F_2$ | 6.8281 | $3d^3\ 4s^2\ {}^4F$ | 0.08 (14) |
| 23 | V $3d^3\ 4s^2\ {}^4F_{3/2}$ | 6.7462 | $3d^4\ 4s^2\ {}^5D$ | 0.526 (12) |
| 24 | Cr $3d^5\ 4s\ {}^7S_3$ | 6.7665 | $3d^5\ 4s^2\ {}^6S_{3/2}$ | 0.667 (10) |
| 25 | Mn $3d^5 4s^2\ {}^6S_{5/2}$ | 7.434 | $3d^6\ 4s^2\ {}^5D$ | < 0 |
| 26 | Fe $3d^6 4s^2\ {}^5D_4$ | 7.9024 | $3d^7 4s^2\ {}^4F$ | 0.164 (35) |
| 27 | Co $3d^7 4s^2\ {}^4F_{9/2}$ | 7.881 | $3d^8\ 4s^2\ {}^3F$ | 0.662 (10) |
| 28 | Ni $3d^8\ 4s^2\ {}^3F_4$ | 7.6398 | $3d^9 4s^2\ {}^2D_{5/2}$ | 1.15 (10) |
| 29 | Cu $3d^{10}\ 4s\ {}^2S_{1/2}$ | 7.7264 | $3d^1 4s^2\ {}^1S_0$ | 1.226 (10) |
| 30 | Zn $4s^2\ {}^1S_0$ | 9.9342 | $4s^2\ 5s\ {}^2S_{1/2}$ ( ?) | ~0 |
| 31 | Ga $4p\ {}^3P_{1/2}$ | 5.9993 | $4p^2\ {}^3P_0$ | 0.30 (15) |
| 32 | Ge $4p^2\ {}^3P_0$ | 7.8994 | $4p^3\ {}^4S_{3/2}$ | 1.2 (1) |
| | Ge $4p^2\ {}^3P_0$ | | $4p^3\ {}^2D$ | ~0.4 |
| 33 | As $4p^3\ {}^4S_{3/2}$ | 9.7886 | $4p^4\ {}^3P_2$ | 0.80 (5) |
| 34 | Se $4p^4\ {}^3P_2$ | 9.7524 | $4p^5\ {}^2P_{3/2}$ | 2.0206 (3) |
| 35 | Br $4p^5\ {}^2P_{3/2}$ | 11.8138 | $4p^6\ {}^1S_0$ | 3.363590(3)[4] |
| 36 | Kr $4p^6\ {}^1S_0$ | 13.9996 | | < 0 |
| 37 | Rb $5s\ {}^2S_{1/2}$ | 4.1771 | $5s^2\ {}^1S_0$ | 0.48592 (2) |
| 38 | Sr $5s^2\ {}^1S_0$ | 5.6949 | $4d\ 5s^2\ {}^2D$ | < 0 |
| 39 | Y $4d\ 5s^2\ {}^2D_{3/2}$ | 6.2171 | $4d\ 5s^2\ 5p\ {}^1D_2$ | 0.307 (12) |
| | | | $4d\ 5s^2\ 5p\ {}^3D_1$ | 0.164 (25) |

| Z | Atom | Ionisation energy | Negative Ion | Electron affini |
|---|---|---|---|---|
| 40 | Zr $4d^2\ 5s^2\ {}^3F_2$ | 6.6339 | $4d^3\ 5s^2\ {}^4F$ | 0.427 (14) |
| 41 | Nb $4d^4\ 5s\ {}^6D_{1/2}$ | 6.7589 | $4d^4\ 5s^2\ {}^5D$ | 0.894 (25) |
| 42 | Mo $4d^5\ 5s\ {}^7S_3$ | 7.0924 | $4d^5\ 5s^2\ {}^6S_{5/2}$ | 0.747 (10) |
| 43 | Tc $4d^5\ 5s^2\ {}^6S_{5/2}$ | 7.28 | $4d^6\ 5s^2\ {}^5D$ | 0.55 (20) |
| 44 | Ru $4d^7\ 5s\ {}^5F_5$ | 7.3605 | $4d^7\ 5s^2\ {}^4F$ | 1.05 (15) |
| 45 | Rh $4d^8\ 5s\ {}^4F_{9/2}$ | 7.4589 | $4d^8\ 5s^2\ {}^3F$ | 1.138 (8) |
| 46 | Pd $4d^{10}\ {}^1S_0$ | 8.3369 | $4d^9\ 5s^2\ {}^2D_{5/2}$ | 0.136 (8) |
| | Pd $4d^{10}\ {}^1S_0$ | | $4d^9\ 5s^2\ {}^2S_{1/2}$ | 0.558 (8) |
| 47 | Ag $4d^{10}\ 5s\ {}^2S_{1/2}$ | 7.5762 | $4d^{10}\ 5s^2\ {}^1S_0$ | 1.303 (7) |
| 48 | Cd $4d^{10}\ 5s^2\ {}^1S_0$ | 8.9938 | $5s^2\ 6s\ {}^2S_{1/2}$ | ≤ 0 |
| | Cd $4d^{10}\ 5s^2\ {}^1S_0$ | | $5s^2\ 6p\ {}^2P$ | < 0 |
| 49 | In $5p\ {}^2P_{1/2}$ | 5.7864 | $5p^2\ {}^3P_0$ | 0.30 (15) |
| 50 | Sn $5p^2\ {}^3P_0$ | 7.3439 | $5p^3\ {}^4S_{3/2}$ | 1.25 (10) |
| | Sn $5p^2\ {}^3P_0$ | | $5p^3\ {}^2D$ (m) | 0.4 (2) |
| 51 | Sb $5p^3\ {}^4S_{3/2}$ | 8.6084 | $5p^4\ {}^3P_3$ | 1.05 (5) |
| 52 | Te $5p^4\ {}^3P_2$ | 9.0096 | $5p^5\ {}^2P_{3/2}$ | 1.9708 (3) |
| 53 | I $5p^5\ {}^2P_{3/2}$ | 10.4513 | $5p^6\ {}^1S_0$ | 3.0591 (4) |
| 54 | Kr $5p^6\ {}^1S_0$ | 12.1298 | | <0 |
| 55 | Cs $6s\ {}^2S_{1/2}$ | 3.8939 | $6s^2\ {}^1S_0$ | 0.4715 |
| 56 | Ba $6s^2\ {}^1S_0$ | 5.2117 | $5d\ 6s^2\ {}^2D$ | < 0 |
| 57 | La $5d\ 6s^2\ {}^2D_{3/2}$ | 5.5769 | $5d^2\ 6s^2\ {}^3F_2$ | 0.5 (3) |
| | Rare Earths | | | ≤1.0 |
| 72 | Hf $5d^2\ 6s^2\ {}^3F_2$ | 6.8251 | $5d^3\ 6s^2\ {}^4F$ | ? |
| 73 | Ta $5d^3\ 6s^2\ {}^4F_{3/2}$ | 7.5496 | $5d^4\ 6s^2\ {}^5D$ | 0.323 (12) |
| 74 | W $5d^4\ 6s^2\ {}^5D_0$ | 7.864 | $5d^5\ 6s^2\ {}^6S_{5/2}$ | 0.816 (8) |
| 75 | Re $5d^5\ 6s^2\ {}^6S_{3/2}$ | 7.8335 | $5d^6\ 6s^2\ {}^5D$ | 0.15 (10) |
| 76 | Os $5d^6\ 6s^2\ {}^5D_4$ | 8.4382 | $5d^7\ 6s^2\ {}^4F$ | 1.1 (3) |
| 77 | Ir $5d^7\ 6s^2\ {}^4F_{9/2}$ | 8.967 | $5d^8\ 6s^2\ {}^3F$ | 1.566 (8) |
| 78 | Pt $5d^9\ 6s\ {}^3D_3$ | 8.9587 | $5d^9\ 6s^2\ {}^2D_{5/2}$ | 2.128 |

| Z | Atom | Ionisation energy | Negative Ion | Electron affini |
|---|---|---|---|---|
| 79 | Au $5d^{10}$ $6s$ $^2S_{1/2}$ | 9.2255 | $5d^{10}$ $6s^2$ $^1S_0$ | 2.3086 (7) |
| 80 | Hg $6s^2$ $^1S_0$ | 10.4375 | $6s^2$ $7s$ $^2S_{1/2}$ | < 0 |
| | Hg $6s^2$ $^1S_0$ | | $6s^2$ $6p$ $^2P_{1/2}$ | <0 |
| 81 | Tl $6s^2 6p$ $^2P_{1/2}$ | 6.1082 | $6p^2$ $^3P_0$ | 0.3 (2) |
| 82 | Pb $6p^2$ $^3P_0$ | 7.4167 | $6p^3$ $^4S_{3/2}$ | 0.365 (8) |
| | Pb $6p^2$ $^3P_0$ | | $6p^3$ $^2D_{3/2}$ | > 0 ? |
| 83 | Bi $6p^3$ $^4S_{3/2}$ | 7.2856 | $6p^4$ $^3P_2$ | 0.947 (10) |
| 84 | Po $6p^4$ $^3P_2$ | 8.417 | $6p^5$ $^2P_{3/2}$ | 1.9 (3) |
| 85 | At $6p^5$ $^2P_{3/2}$ | | $6p^6$ $^1S_0$ | 2.8 (2) |
| 86 | Rn $6p^6$ $^1S_0$ | 10.7485 | | < 0 |